\documentclass[10pt]{article}
\usepackage{amsfonts,latexsym}
\usepackage{amsmath}
\usepackage{amscd}
\usepackage{float,amsmath,amssymb,mathrsfs,bm,multirow,graphics}
\usepackage[dvips]{graphicx}
\usepackage[round]{natbib}

\newcommand{\nequation}{\setcounter{equation}{0}}

\newcommand{\R}{{\Bbb R}}

\newcommand{\C}{{\Bbb C}}

\newcommand{\proofbegin}{\noindent{\it Proof.\,\,}}
\newcommand{\proofend}{\hfill$\Box$\bigskip}

\newtheorem{theorem}{Theorem}[section]
\newtheorem{proposition}[theorem]{Proposition}

\newtheorem{figuretext}{Figure}

\input epsf
\title{\sc An integrable generalization of the sine-Gordon equation on the half-line}
\author{Jonatan Lenells}
\date{{\small Department of Mathematics, Baylor University, One Bear Place 97328, \\ Waco, TX 76798, USA. \\ 
E-mail: Jonatan\_Lenells@baylor.edu
}}

\begin{document}
\maketitle
\begin{abstract} 
\noindent
We analyze a generalization of the sine-Gordon equation in laboratory coordinates on the half-line. Using the Fokas transform method for the analysis of initial-boundary value problems for integrable PDEs, we show that the solution $u(x,t)$ can be constructed from the initial and boundary values via the solution of a $2\times 2$-matrix Riemann-Hilbert problem.
\end{abstract}




\section{Introduction} \nequation
We consider the following integrable generalization of the sine-Gordon equation:
\begin{equation}\label{gsglab}
  u_{xx} - u_{tt} = \left(1 + \nu (\partial_x + \partial_t)^2\right)\sin{u(x,t)}, \qquad x, t \in \R,
\end{equation}
where $u(x,t)$ is a real-valued function and $\nu \in \R$ is a parameter---note that (\ref{gsglab}) reduces to the sine-Gordon equation in laboratory coordinates when $\nu = 0$. 
In terms of the `light-cone' coordinates $(\xi, \eta)$ defined by 
\begin{equation}\label{lighttolab}
  x = \xi + \eta, \qquad t = \xi - \eta,
\end{equation} 
equation (\ref{gsglab}) takes the form
\begin{equation}\label{gsg}
  u_{\xi \eta} = (1 + \nu \partial_\xi^2)\sin(u), \qquad \xi, \eta \in \R.
\end{equation}
Equation (\ref{gsg}) was derived using bi-Hamiltonian methods in \citet{F1995}. It is related to the sine-Gordon equation in the same way that the Camassa-Holm equation is related to the KdV equation.

In this paper, we will assume that $\nu < 0$ and for simplicity set $\nu = -1$. For this value of $\nu$, equation (\ref{gsg}) appeared in \citet{SaSa}, where it was shown to be related, via certain transformations, to an integrable equation which describes pseudospherical surfaces introduced in \citet{Ra}. The inverse scattering transform (IST) formalism on the line for equation (\ref{gsg}) with $\nu = -1$ was implemented in \citet{LFgsg}.

\begin{figure}
\begin{center}
    \includegraphics[width=.5\textwidth]{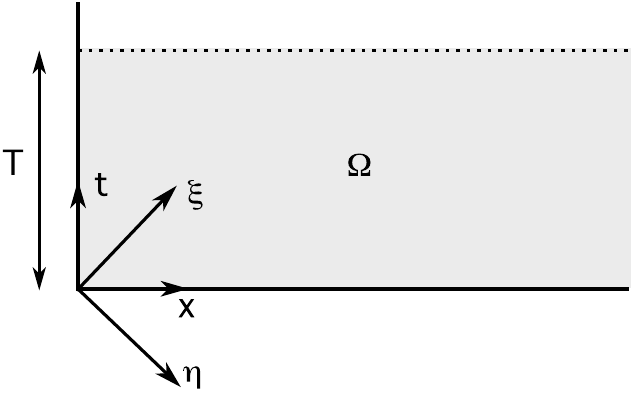}
     \begin{figuretext}\label{halflinedomain.pdf}
       The half-line domain $\Omega$ with respect to the laboratory and light-cone coordinates.
     \end{figuretext}
     \end{center}
\end{figure}

A method for the analysis of initial-boundary value (IBV) problems for nonlinear integrable PDEs was announced in \citet{F1997} and subsequently developed further by several authors, see \citet{Fbook}. Here, we use this method to study equation (\ref{gsglab}) in the half-line domain
$$\Omega = \{(x,t) \, |\, 0 \leq x < \infty,\, 0 \leq t \leq T\},$$
where $T \leq \infty$ is a given final time, see Figure \ref{halflinedomain.pdf}. Given initial values at $x = 0$ and boundary values at $t = 0$ such that the corresponding IBV problem for (\ref{gsglab}) in the domain $\Omega$ has a solution $u(x,t)$, we show that $u(x,t)$ can be constructed via the solution of a $2\times 2$-matrix Riemann-Hilbert (RH) problem. The main notable features as compared with other similar applications of the methodology of \citet{F1997} are: (1) The formulation of the RH problem suggested by the above methodology depends, in addition to the variables $(x,t)$, on a function $p(x,t)$ which is unknown from the point of view of the inverse problem. In order to formulate a RH problem whose jump matrix involves only known quantities, we have to reparametrize the $x$ and $t$ variables. A similar situation occurs in the analysis of the half-line problem for the Camassa-Holm equation, where however only the $x$-variable has to be reparametrized, see \citet{BS2008}. (2) Only certain combinations of $u$ and its derivatives can be recovered from the RH problem. Therefore, in addition to solving the RH problem, the reconstruction of $u(x,t)$ involves finding the solution of a nonlinear ODE. Following the ideas of \citet{LFgsg} we show that this ODE can be reduced to a Ricatti equation. 
(3) The adopted Lax pair for equation (\ref{gsglab}) has singularities at $\lambda = \infty$ and $\lambda = 0$, where $\lambda \in \hat{\C} = \C \cup \{\infty\}$ denotes the spectral parameter. In order to define eigenfunctions which are bounded on the whole Riemann $\lambda$-sphere, we will use two different representations of the Lax pair. These representations are suitable for the definition of eigenfunctions which are bounded near $\lambda = \infty$ and $\lambda = 0$, respectively.

The analogous problem for the sine-Gordon equation on the half-line (i.e. for the equation obtained from (\ref{gsglab}) by letting $\nu = 0$) was investigated in \citet{F2004, Fbook, FI92}. We emphasize that although there exists (when $\nu < 0$) a Liouville type transformation relating equation (\ref{gsglab}) to the sine-Gordon equation (see \citet{SaSa, LFgsg}), the half-line problems for these two equations are not equivalent, since the Liouville transformation transformation distorts the shape of the domain $\Omega$. 

In section \ref{eigensec} we introduce a Lax pair for equation (\ref{gsglab}) and define bounded and analytic eigenfunctions which are suitable for the formulation of a RH problem. The jump matrix of this RH problem can be expressed in terms of certain spectral functions, which are introduced in section \ref{specsec}. Finally, the main result is stated in section \ref{RHsec}.

\section{Bounded and analytic eigenfunctions}\label{eigensec} \nequation
The Riemann-Hilbert formalism for integrating a nonlinear evolution equation is based on the construction of eigenfunctions of the associated Lax pair. These eigenfunctions are joined together to a bounded and sectionally analytic function on the Riemann sphere of the spectral parameter $\lambda \in \hat{\C} = \C \cup \{\infty\}$. A Lax pair suitable for the construction of eigenfunctions which are bounded near $\lambda = \infty$ was derived in \citet{LFgsg}. For the problem on the line, this Lax pair representation alone was sufficient for the formulation of a RH problem, since the eigenfunctions could be constructed using only the $x$-part of the Lax pair. For the problem on the half-line, the construction of eigenfunctions involves also the $t$-part of the Lax pair, which has a singularity at $\lambda = 0$. We will therefore introduce another representation of the Lax pair suitable for the construction of eigenfunctions which are bounded near $\lambda = 0$. Then, according to the methodology of \citet{F1997}, we will define solutions of these Lax pair representations by integration from three different corners of the spatial domain $\Omega$. The eigenfunctions which are bounded near $\lambda = 0$ and $\lambda = \infty$ will be denoted by $\{\mu_j\}_1^3$ and $\{\Phi_j\}_2^3$, respectively. Together the $\mu_j$'s and the $\Phi_j$'s can be used to formulate a $2 \times 2$-matrix RH problem.

\subsection{Lax pair representations}
Let
$$\sigma_1 = \begin{pmatrix} 0	&	1 \\ 1	& 0\end{pmatrix}, \qquad
\sigma_2 = \begin{pmatrix} 0	&	-i \\ i	& 0\end{pmatrix}, \qquad
\sigma_3 = \begin{pmatrix} 1	&	0 \\ 0	& -1\end{pmatrix},$$
and define $m(x,t)$ by
\begin{equation}\label{mdef}
  m(x,t) = 1 + (u_x(x,t) + u_t(x,t))^2.
\end{equation} 
Applying the change of variables (\ref{lighttolab}) to the Lax pair of equation (\ref{gsg}) derived in \citet{LFgsg}, we find the following Lax pair for equation (\ref{gsglab}):
\begin{equation}\label{lax}
\begin{cases}
	& \phi_x + i\left( \lambda p_x  - \frac{1}{8\lambda}\right)\sigma_3 \phi = W_+ \phi,	\\
	& \phi_t  + i\left( \lambda p_t  + \frac{1}{8\lambda}\right)\sigma_3 \phi = W_- \phi,
\end{cases}
\end{equation}	
where $\phi(x,t,\lambda)$ is a $2 \times 2$-matrix valued eigenfunction, $\lambda \in \hat{\C} = \C \cup \{\infty\}$ is a spectral parameter, the functions $W_\pm(x,t,\lambda)$ are defined by
\begin{align}\label{Wpmdef}
W_\pm = & \mp \frac{i}{8\lambda}\sigma_3 \pm i\frac{\cos(u) - (u_t + u_x)\sin(u)}{8\sqrt{m}\lambda} \sigma_3 \pm i \frac{(u_t + u_x)\cos(u) + \sin(u)}{8\sqrt{m}\lambda} \sigma_2
	\\ \nonumber
&+ \frac{i(1 + \cos{u})(m(u_t + u_x)(1 - \cos{u}) - m \sin(u) + 2u_{xx} + 2u_{xt})}{4(1 \pm \cos{u}) m}\sigma_1.
\end{align}
and $p(x,t)$ is a real-valued function such that
\begin{equation}\label{pxpt}
  p_x = \frac{1}{2}(1 - \cos{u})\sqrt{m}, \qquad p_t = \frac{1}{2}(1 + \cos{u})\sqrt{m}.
\end{equation}
The equations in (\ref{pxpt}) are compatible since equation (\ref{gsglab}) admits the conservation law
\begin{equation}\label{conslaw}
  \left((1 - \cos{u})\sqrt{m}\right)_t = \left((1 + \cos{u})\sqrt{m}\right)_x.
\end{equation}
We choose $p(x,t)$ so that $p(0,0) = 0$, i.e.
\begin{equation}\label{pdef}
  p(x,t) = \frac{1}{2} \int_{(0,0)}^{(x,t)} \left[(1 - \cos{u})\sqrt{m}dx' + (1 + \cos{u})\sqrt{m}dt'\right].
\end{equation}
Despite the form of the denominator of the last term in (\ref{Wpmdef}), the functions $W_\pm$ do not have singularities at points where $1 \pm \cos{u} = 0$. Indeed, using equation (\ref{gsglab}), the last term on the right-hand of (\ref{Wpmdef}) can be rewritten as
$$ i(1 \mp \cos{u})\frac{m(u_t + u_x) + u_{tt} + 2u_{tx} + u_{xx}}{4m}\sigma_1,$$
and this expression is manifestly nonsingular.

The functions $W_\pm$ have the following properties:
\begin{itemize}
\item $W_\pm(x, t, \lambda) \to 0, \qquad x \to \infty$,
\item $W_\pm(x, t,\lambda) = \begin{pmatrix} O(1/\lambda) & O(1) \\ O(1) & O(1/\lambda) \end{pmatrix}, \qquad \lambda \to \infty,$
\item $\text{tr}(W_\pm(x,t,\lambda)) = 0,$
\item $W_\pm^\dagger(x,t,\bar{\lambda}) = -W_\pm(x,t,\lambda)$,
\end{itemize}
where $A^\dagger$ denotes the complex-conjugate transpose of a matrix $A$. The last two of these properties ensure that the eigenfunction $\phi(x,t,\lambda)$ can be normalized so that
\begin{equation}\label{phisymm}
\det(\phi(x,t, \lambda)) = 1, \qquad 
\phi^\dagger(x,t,\bar{\lambda}) = \phi^{-1}(x,t,\lambda).
\end{equation}

The Lax pair (\ref{lax}) is convenient for the definition of eigenfunctions which are bounded near $\lambda = \infty$. 
In order to define eigenfunctions which are bounded near $\lambda = 0$, we transform the Lax pair as follows. Let $I$ denote the $2 \times 2$ identity matrix. The gauge transformation
\begin{equation}\label{phigpsi}
  \phi(x,t,\lambda) = g(x,t) \psi(x,t,\lambda),
\end{equation}  
where
\begin{align}\label{gdef}
  g(x,t) = \frac{1}{\sqrt{2}(u_x + u_t)}&\sqrt{1 + \frac{1}{\sqrt{m}}}\biggl[\left((\sqrt{m} - 1) \cos(u/2) + (u_x + u_t)\sin(u/2)\right) I 
  	\\ \nonumber
  &+ i\left(-(u_x + u_t)\cos(u/2) + (\sqrt{m} - 1)\sin(u/2)\right)\sigma_1\biggr],
\end{align}
transforms (\ref{lax}) into 
\begin{equation}\label{laxzero}
\begin{cases}
	& \psi_x + \frac{i}{8\lambda} \sigma_3 \psi = V_1 \psi,	\\
	& \psi_t - i \left(\lambda + \frac{1}{8\lambda}\right) \sigma_3 \psi = V_2 \psi,
\end{cases}
\end{equation}	
where $V_j = V_j(x,t,\lambda)$, $j = 1,2$, are defined by
\begin{align} \nonumber
V_1 = \; & i\sin^2(u/2)(\cos(u) - (u_t + u_x)\sin{u}) \lambda \sigma_3
	\\ \nonumber
 & - i \sin^2(u/2)((u_t + u_x)\cos(u) + \sin{u})\lambda \sigma_2
	\\ \label{V1V2def}
& -\frac{i}{4}\left(\sin(u) - u_t(1 - \cos{u}) + u_x (1 + \cos{u})\right) \sigma_1,
	\\ \nonumber
V_2 = \; & \frac{i}{4}(-3 + 2\cos(u) + \cos(2u) - (u_t + u_x)(2\sin(u) + \sin(2u)))\lambda \sigma_3 
	\\ \nonumber
& - i \cos^2(u/2)((u_t + u_x)\cos(u) + \sin{u})\lambda \sigma_2
	\\ \nonumber
& + \frac{i}{4}\left(\sin(u) - u_t(1 - \cos{u}) + u_x (1 + \cos{u})\right)\sigma_1. 
\end{align}
The form (\ref{gdef}) of $g$ is motivated by the fact that it diagonalizes the terms of highest order as $\lambda \to 0$ of the Lax pair (\ref{lax}) and that it satisfies
\begin{equation}\label{gsymm}
  \det(g(x,t)) = 1, \qquad g^\dagger(x,t) = g^{-1}(x,t).
\end{equation}
The relations (\ref{gsymm}) ensure that the gauge transformation (\ref{phigpsi}) preserves the properties in (\ref{phisymm}), i.e.
\begin{equation}\label{psisymm}
\det(\psi(x,t, \lambda)) = 1, \qquad 
\psi^\dagger(x,t,\bar{\lambda}) = \psi^{-1}(x,t,\lambda).
\end{equation}
The function $g(x,t)$ is nonsingular as $u_x + u_t \to 0$ despite the form of the right-hand side of (\ref{gdef}). In fact,
$$g(x,t) \to \begin{pmatrix} \sin(u/2) & -i\cos(u/2) \\ -i\cos(u/2)  & \sin(u/2)\end{pmatrix}\quad \text{as} \quad u_x + u_t \to 0.$$
The functions $V_1$ and $V_2$ have the following properties:
\begin{itemize}
\item $V_j(x, t, \lambda) \to 0, \qquad x \to \infty, \quad j = 1,2$,
\item $V_j(x, t,\lambda) = \begin{pmatrix} O(\lambda) & O(1) \\ O(1) & O(\lambda) \end{pmatrix}, \qquad \lambda \to 0, \quad j = 1,2,$
\item $\text{tr}(V_j(x,t,\lambda)) = 0, \qquad j = 1,2,$
\item $V_j^\dagger(x,t,\bar{\lambda}) = -V_j(x,t,\lambda), \quad j = 1,2$.
\end{itemize}

\subsection{Eigenfunctions bounded near $\lambda = 0$}\label{eigenzerosubsec}
In this subsection, we define solutions of (\ref{laxzero}) which are well-behaved near $\lambda = 0$. 
Introducing an eigenfunction $\mu$ by
\begin{equation}\label{psimurel}
  \psi = \mu e^{-i\left(\frac{x}{8\lambda} - (\lambda + \frac{1}{8\lambda}) t\right) \sigma_3},
\end{equation}
we find that the Lax pair (\ref{laxzero}) becomes
\begin{equation}\label{mulax}
\begin{cases}
	& \mu_x + \frac{i}{8\lambda} [\sigma_3, \mu] = V_1 \mu,	\\
	& \mu_t - i \left(\lambda + \frac{1}{8\lambda}\right) [\sigma_3, \mu] = V_2 \mu.
\end{cases}
\end{equation}	
\begin{figure}
\begin{center}
    \includegraphics[width=.3\textwidth]{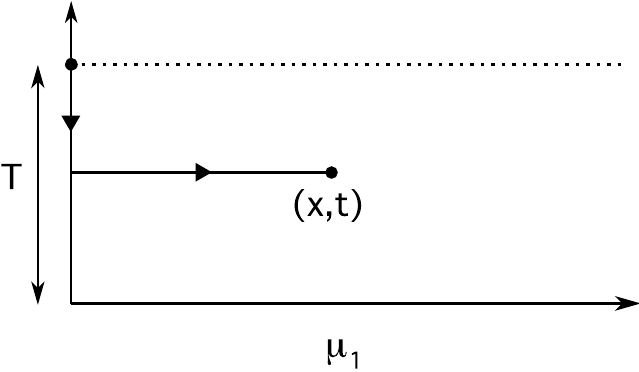} \quad
   \includegraphics[width=.3\textwidth]{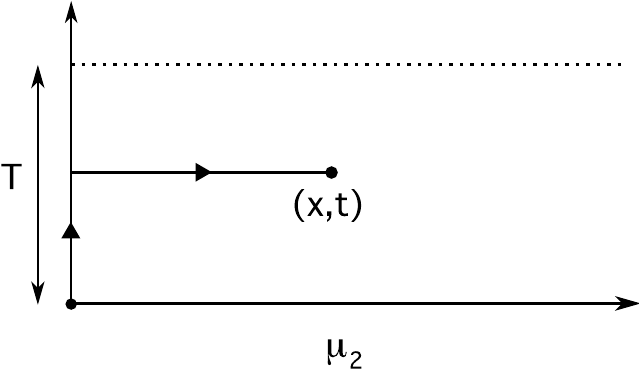} \quad
   \includegraphics[width=.3\textwidth]{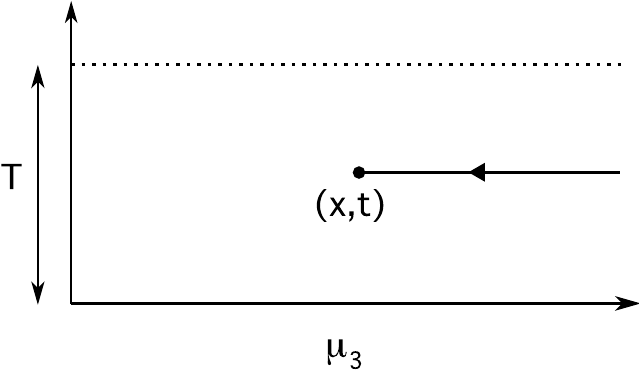}  \\
     \begin{figuretext}\label{mu123.pdf}
       The contours of integration for the solutions $\mu_1$, $\mu_2$, and $\mu_3$ of (\ref{mulax}).
     \end{figuretext}
     \end{center}
\end{figure}
This can be written in differential form as
\begin{equation}\label{laxzerodiff}
  d\left(e^{i \left(\frac{x}{8\lambda} - (\lambda + \frac{1}{8\lambda}) t\right) \hat{\sigma}_3} \mu\right) = W,
\end{equation}
where $\hat{\sigma}_3$ acts on a $2\times 2$ matrix $A$ by $\hat{\sigma}_3A = [\sigma_3, A]$, and the closed one-form $W(x,t,\lambda)$ is defined by
\begin{equation}\label{Wzerodef}
  W = e^{i \left(\frac{x}{8\lambda} - (\lambda + \frac{1}{8\lambda}) t\right) \hat{\sigma}_3} (V_1 dx + V_2 dt)\mu.
\end{equation}
We define three eigenfunctions $\{\mu_j\}_1^3$ of (\ref{laxzerodiff}) by
\begin{equation}\label{mujdef}
   \mu_j(x,t,\lambda) = I + \int_{(x_j, t_j)}^{(x,t)} e^{-i \left(\frac{x}{8\lambda} - (\lambda + \frac{1}{8\lambda}) t\right) \hat{\sigma}_3}W(x',t',\lambda),
\end{equation}
where $(x_1, t_1) = (0, T)$, $(x_2, t_2) = (0, 0)$, and $(x_3, t_3) = (\infty, t)$. Since the one-form $W$ is exact, the integral on the right-hand side of (\ref{mujdef}) is independent of the path of integration. We choose the particular contours shown in Figure \ref{mu123.pdf}. 
This choice implies the following relations on the contours:
\begin{align} \nonumber
(x_1, t_1) \to (x,t): x' - x \leq 0,& \qquad t' - t \geq 0,
	\\ \label{contourrelations}
(x_2, t_2) \to (x,t): x' - x \leq 0,& \qquad t' - t \leq 0,
	\\ \nonumber
(x_3, t_3) \to (x,t): x' - x \geq 0, & \qquad t' - t = 0.
\end{align}
\begin{figure}
\begin{center}
    \includegraphics[width=.4\textwidth]{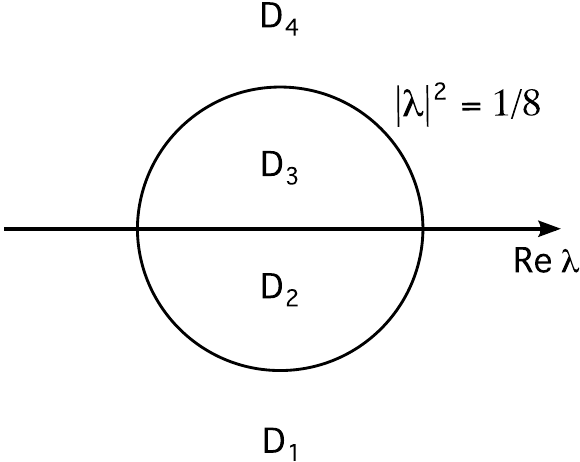} \quad  \\
     \begin{figuretext}\label{D1234.pdf}
       The sets $\{D_j\}_1^4$ in the complex $\lambda$-plane.
     \end{figuretext}
     \end{center}
\end{figure}
Letting
\begin{align}\nonumber
&D_1 = \left\{\lambda \in \hat{\C} \biggl| \text{Im}\, \frac{1}{8\lambda} > 0  \text{  and  } \text{Im}\left(\lambda + \frac{1}{8\lambda}\right) < 0 \right\},
	\\ \label{D1234def}
&D_2 = \left\{ \lambda \in \hat{\C} \biggl| \text{Im}\, \frac{1}{8\lambda} > 0  \text{  and  }  \text{Im}\left(\lambda +  \frac{1}{8\lambda}\right) > 0 \right\},
	\\ \nonumber
&D_3 = \left\{\lambda \in \hat{\C} \biggl|\text{Im}\, \frac{1}{8\lambda} < 0  \text{  and  }  \text{Im}\left(\lambda + \frac{1}{8\lambda}\right) < 0 \right\},
	\\ \nonumber
&D_4 = \left\{ \lambda \in \hat{\C} \biggl| \text{Im}\, \frac{1}{8\lambda} < 0  \text{  and  }  \text{Im}\left(\lambda + \frac{1}{8\lambda}\right) > 0 \right\},
\end{align}
the second column vectors of $\mu_1$, $\mu_2$, $\mu_3$ are analytic for $\lambda \in \hat{\C}$ such that $\lambda$ belongs to $D_3$, $D_4$, and $D_1 \cup D_2$, respectively, see Figure \ref{D1234.pdf}. Moreover, away from $\lambda = \infty$ where the Lax pair is singular, they have continuous and bounded extensions to the closures of these sets.
We will denote these vectors with the superscripts $(3)$, $(4)$, and $(12)$ to indicate these boundedness properties.
Similar conditions are valid for the first column vectors. We obtain
$$\mu_1 = \left(\mu_1^{(2)}, \mu_1^{(3)}\right), \quad 
\mu_2 = \left( \mu_2^{(1)}, \mu_2^{(4)}\right), \quad 
\mu_3 = \left( \mu_3^{(34)}, \mu_3^{(12)}\right).$$
The $\mu_j$'s satisfy
\begin{equation}\label{mujsymm}
\det(\mu_j(x,t, \lambda)) = 1, \qquad 
\mu_j(x,t,\bar{\lambda})^\dagger = \mu_j(x,t,\lambda)^{-1}, \qquad j = 1,2,3.
\end{equation}
Moreover, as $\lambda \to 0$,
\begin{align}
\left(\mu_1^{(2)}(x,t,\lambda), \mu_3^{(12)}(x,t,\lambda)\right) = I + O(\lambda), \qquad \lambda \to 0, \quad \lambda \in D_2,
	\\ \nonumber
\left(\mu_3^{(34)}(x,t, \lambda), \mu_1^{(3)}(x,t, \lambda)\right) = I + O(\lambda), \qquad \lambda \to 0, \quad \lambda \in D_3.
\end{align}
The $\mu_j$'s are suitable for the formulation of a RH problem except that they have singularities at $\lambda = \infty$. Our strategy is therefore to cut out a neighborhood of $\lambda = \infty$ and use the Lax pair (\ref{lax}) to define eigenfunctions which are bounded in this neighborhood.

\subsection{Eigenfunctions bounded near $\lambda = \infty$}\label{eigeninftysubsec}
The form of the Lax pair (\ref{lax}) is convenient for the definition of eigenfunctions which are well-behaved near $\lambda = \infty$. 
Introducing an eigenfunction $\Phi$ by
\begin{equation}\label{phiPhirel}
  \phi = \Phi e^{-i (\lambda p + \frac{t - x}{8\lambda}) \sigma_3},
\end{equation}  
we find that the Lax pair (\ref{lax}) becomes
\begin{equation}\label{Philax}
\begin{cases}
	& \Phi_x + i\left( \lambda p_x  - \frac{1}{8\lambda}\right)[\sigma_3, \Phi] = W_+ \Phi,	\\
	& \Phi_t  + i\left( \lambda p_t  + \frac{1}{8\lambda}\right)[\sigma_3, \Phi] = W_- \Phi.
\end{cases}
\end{equation}	
This can be written in differential form as
\begin{equation}\label{Philaxdiff}
  d\left(e^{i (\lambda p + \frac{t - x}{8\lambda}) \hat{\sigma}_3} \Phi\right) = W_\infty,
\end{equation}
where the closed one-form $W_\infty(x,t, \lambda)$ is defined by
\begin{equation}\label{Winftydef}
  W_\infty = e^{i (\lambda p + \frac{t - x}{8\lambda}) \hat{\sigma}_3} (W_+ dx + W_- dt)\Phi.
\end{equation}
We define two eigenfunctions $\Phi_2$ and $\Phi_3$ of (\ref{Philaxdiff}) by
\begin{equation}\label{Phijdef}
   \Phi_j(x,t, \lambda) = I + \int_{(x_j, t_j)}^{(x,t)} e^{-i (\lambda p(x,t) + \frac{t - x}{8\lambda}) \hat{\sigma}_3}W_\infty(x',t',\lambda),
\end{equation}
where $(x_2, t_2) = (0, 0)$ and $(x_3, t_3) = (\infty, t)$. The functions $\Phi_2$ and $\Phi_3$ are the analogs of $\mu_2$ andÊ $\mu_3$ defined in (\ref{mujdef}); the analog of $\mu_1$ is not needed since we only consider a neighborhood of $\lambda = \infty$. Choosing the integration contours in Figure \ref{mu123.pdf}, the integral equations (\ref{Phijdef}) defining $\Phi_2$ and $\Phi_3$ become
\begin{align*}
   \Phi_2(x,t, \lambda) =\; & I + \int_{0}^{t} e^{i \left(\lambda (p(0,t') - p(x,t)) + \frac{t' - t + x}{8\lambda}\right) \hat{\sigma}_3} (W_- \Phi_2)(0,t', \lambda) dt'
   	\\
& + \int_{0}^{x} e^{i \left(\lambda (p(x',t) - p(x,t)) - \frac{x' - x}{8\lambda}\right) \hat{\sigma}_3} (W_+ \Phi_2)(x',t, \lambda) dx',
   	\\
   \Phi_3(x,t, \lambda) = \; &I - \int_x^\infty e^{i \left(\lambda (p(x',t) - p(x,t)) - \frac{x' - x}{8\lambda}\right) \hat{\sigma}_3} (W_+ \Phi_3)(x',t, \lambda) dx'.
\end{align*}
The second column of the integral equation for $\Phi_2$ involves the exponentials
\begin{equation}\label{mu22exps}
  e^{2i \left(\lambda (p(0,t') - p(x,t)) + \frac{t' - t + x}{8\lambda}\right)} \quad \text{and}\quad e^{2i \left(\lambda (p(x',t) - p(x,t)) - \frac{x' - x}{8\lambda}\right)},
\end{equation}
where the functions $p(\cdot,t)$ and $p(x,\cdot)$ are nondecreasing. Define $R > 0$ by
\begin{equation}\label{Rdef}
  R^2 = \sup_{0 \leq t', t \leq T} \frac{t - t'}{4\int_{t'}^{t} (1 + \cos(u(0, \tau))\sqrt{m(0,\tau)} d\tau}.
\end{equation}
We will henceforth assume that
\begin{equation}\label{cosuass}
  \cos(u(0,t)) \neq -1, \qquad 0 \leq t \leq T,
\end{equation}
so that $R$ is finite. The relations (\ref{contourrelations}) together with the definition of $R$ implies the following inequalities on the contour $(x_2, t_2) \to (x,t)$:
\begin{align} \nonumber
&\text{Im}\left(\lambda (p(0,t') - p(x,t)) + \frac{t' - t + x}{8\lambda}\right) \geq 0 \quad \text{if} \quad |\lambda| \geq R, \quad \text{Im}\,\lambda \leq 0,
	\\ \label{ineqImlambdanegative}
 &\text{Im}\left(\lambda (p(x',t) - p(x,t)) - \frac{x' - x}{8\lambda}\right) \geq 0 \quad \text{if} \quad \text{Im}\,\lambda \leq 0.
\end{align}
The first of these inequalities is a consequence of the estimates
\begin{align*}
\text{Im}\left(\lambda (p(0,t') - p(x,t)) + \frac{t' - t + x}{8\lambda}\right)
& \geq \text{Im}(\lambda) \left(p(0,t') - p(0,t) - \frac{t' - t}{8|\lambda|^2}\right)
	\\
 \geq \text{Im}(\lambda)(&p(0,t') - p(0,t))\left(1 - \frac{R^2}{|\lambda|^2}\right), \qquad \text{Im}\,\lambda \leq 0,
\end{align*}
which hold on the contour $(x_2, t_2) \to (x,t)$.

We define the sets $D_5$ and $D_6$ by
\begin{align}\label{D56def}
&D_5 = \left\{\lambda \in \hat{\C} \bigl| \text{Im}\, \lambda > 0  \text{  and  } |\lambda| > R \right\},
	\\ \nonumber
&D_6 = \left\{ \lambda \in \hat{\C} \bigl| \text{Im}\, \lambda < 0  \text{  and  }   |\lambda| > R \right\}.
\end{align}
The inequalities (\ref{ineqImlambdanegative}) imply that $[\Phi_2]_2$ is bounded and analytic for $\lambda \in D_6$. Similar considerations apply to the other column vectors of $\Phi_2$ andÊ $\Phi_3$, and we deduce that $\Phi_2$ andÊ $\Phi_3$ have the boundedness properties
$$\Phi_2 = \left( \Phi_2^{(5)}, \Phi_2^{(6)}\right), \quad  \Phi_3 = \left( \Phi_3^{(6)}, \Phi_3^{(5)}\right).$$

Moreover, the functions $\Phi_2$ and $\Phi_3$ satisfy
\begin{equation}\label{Phijsymm}
\det(\Phi_j(x,t, \lambda)) = 1, \qquad 
\Phi_j(x,t,\bar{\lambda})^\dagger = \Phi_j(x,t,\lambda)^{-1}, \qquad j = 2,3.
\end{equation}

\section{Spectral functions}\label{specsec} \nequation
We define two $2\times 2$-matrix valued spectral functions $s(\lambda)$ and $S(\lambda)$ by\begin{subequations}\label{sSdef}
\begin{align}
\label{sdef} 
  \mu_3(x,t,\lambda) &= \mu_2(x,t,\lambda)e^{-i\left(\frac{x}{8\lambda} - (\lambda + \frac{1}{8\lambda}) t\right)\hat{\sigma}_3} s(\lambda),
		\\
  \label{Sdef} 
  \mu_1(x,t,\lambda) &= \mu_2(x,t,\lambda)  e^{-i\left(\frac{x}{8\lambda} - (\lambda + \frac{1}{8\lambda}) t\right)\hat{\sigma}_3} S(\lambda).
\end{align}
\end{subequations}
Evaluation of (\ref{sSdef}) at $(x,t) = (0,0)$ and $(x,t) = (0,T)$ gives the following expressions for $s(\lambda)$ and $S(\lambda)$:
\begin{equation}\label{sSmurel}   
   s(\lambda) = \mu_3(0,0, \lambda), \qquad S(\lambda) = \mu_1(0,0, \lambda) = \left(e^{-i(\lambda + \frac{1}{8\lambda}) T \hat{\sigma}_3}\mu_2(0,T,\lambda)\right)^{-1}.
\end{equation}
The function $\mu_3(x,0, \lambda)$ is defined by the integral equation obtained by setting $t = 0$ in (\ref{mujdef}). This integral equation involves the function $V_1(x,0,\lambda)$ defined in (\ref{V1V2def}).
Thus $s(\lambda)$ is defined in terms of the initial data $u(x,0)$ and $u_t(x,0)$ alone.
Similarly, $\mu_1(0,t,\lambda)$ is defined by the integral equation obtained by setting $x = 0$ in (\ref{mujdef}) which involves $V_2(0,t,\lambda)$. Thus $S(\lambda)$ is defined in terms of the boundary data $u(0,t)$ and $u_x(0,t)$ alone.

We infer from (\ref{mujsymm}) that
\begin{equation}\label{sSdet1}  
  \det s(\lambda) = 1, \qquad \det S(\lambda) = 1,
\end{equation}
and that there exist functions $a(\lambda)$, $b(\lambda)$, $A(\lambda)$, and $B(\lambda)$ such that
$$s(\lambda) = \begin{pmatrix} \overline{a(\bar{\lambda})} & b(\lambda) \\
- \overline{b(\bar{\lambda})} 	&	a(\lambda) \end{pmatrix}, \qquad S(\lambda) = \begin{pmatrix} \overline{A(\bar{\lambda})} & B(\lambda) \\
- \overline{B(\bar{\lambda})} 	&	A(\lambda) \end{pmatrix}.$$
Defining the sets $D_1'$ and $D_4'$ by (see Figure \ref{RHcontour.pdf})
\begin{align}\label{D12primedef}
  D_1' = D_1\setminus\bar{D}_6, \qquad D_4' = D_4\setminus\bar{D}_5,
\end{align}
i.e. $D_1'$ and $D_4'$ denote the sets $D_1$ and $D_4$ with a neighborhood of $\lambda = \infty$ removed, we can state the following result.

\begin{proposition}\label{abABprop}
The spectral functions $a(\lambda)$ and $b(\lambda)$ have the following properties:
\begin{enumerate}
\item[(i)] $a(\lambda)$ and $b(\lambda)$ are analytic for $\lambda \in D_1 \cup D_2$ and continuous and bounded for $\lambda \in \bar{D}'_1 \cup \bar{D}_2$.
\item[(ii)] $a(\lambda) = 1 + O(\lambda), \quad b(\lambda) = O(\lambda), \qquad \lambda \to 0, \quad \lambda \in D_2$.
\item[(iii)] $|a(\lambda)|^2 + |b(\lambda)|^2  = 1, \qquad \lambda \in \R$.
\end{enumerate}
The spectral functions $A(\lambda)$ and $B(\lambda)$ have the following properties:
\begin{enumerate}
\item[(i)] $A(\lambda)$ and $B(\lambda)$ are analytic for $\lambda \in D_1 \cup D_3$ and continuous and bounded for $\lambda \in \bar{D}'_1 \cup \bar{D}_3$. If $T < \infty$, then $A(\lambda)$ and $B(\lambda)$ are defined and analytic in $\hat{\C} \backslash \{0, \infty\}$.
\item[(ii)] $A(\lambda) = 1 + O(\lambda), \quad B(\lambda) = O(\lambda), \qquad \lambda \to 0, \quad \lambda \in D_1 \cup D_3$.
\item[(iii)] $A(\lambda)\overline{A(\bar{\lambda})} + B(\lambda)\overline{B(\bar{\lambda})} = 1$ for $\lambda \in \R \cup \{|\lambda|^2 = 1/8\}$ (for $\lambda \in \C$ if $T < \infty$).
\end{enumerate}
\end{proposition}
\proofbegin
The properties denoted by (i) and (ii) follow from the discussion in subsection \ref{eigenzerosubsec} and the observation that the definition of $\mu_1(0,t,\lambda)$ implies that this function has the following enlarged domain of boundedness:
\begin{equation}\label{mu1largerdomain}
  \mu_1(0,t,\lambda) = \left(\mu_1^{(24)}(0,t,\lambda), \mu_1^{(13)}(0,t,\lambda)\right).
\end{equation}
The properties denoted by (iii) follow from (\ref{sSdet1}).
\proofend

We will also need the spectral function $s_\infty(\lambda)$ defined by
\begin{equation}\label{sinftydef} 
  \Phi_3(x,t,\lambda) = \Phi_2(x,t,\lambda)e^ {-i (\lambda p + \frac{t - x}{8\lambda}) \hat{\sigma}_3} s_\infty(\lambda).
\end{equation}
Evaluation of (\ref{sinftydef}) at $(x,t) = (0,0)$ yields
\begin{equation}\label{sinftyPhirel}
   s_\infty(\lambda) = \Phi_3(0,0, \lambda).
\end{equation}
Just like $s(\lambda)$, the function $s_\infty(\lambda)$ is defined in terms of the initial data $u(x,0)$ and $u_t(x,0)$.
Moreover, $s_\infty$ satisfies 
\begin{equation}\label{sinftydet1}  
  \det s_\infty(\lambda) = 1,
\end{equation}
and can be written as
$$s_\infty(\lambda) = \begin{pmatrix} \overline{a_\infty(\bar{\lambda})} & b_\infty(\lambda) \\
- \overline{b_\infty(\bar{\lambda})} 	&	a_\infty(\lambda) \end{pmatrix}.$$
where $a_\infty(\lambda)$ and $b_\infty(\lambda)$ are complex-valued functions.

\begin{proposition}\label{abinftyprop}
The spectral functions $a_\infty(\lambda)$ and $b_\infty(\lambda)$ have the following properties:
\begin{enumerate}
\item[(i)] $a_\infty(\lambda)$ and $b_\infty(\lambda)$ are analytic in $D_5$ with continuous and bounded extensions to $\lambda \in \bar{D}_5$ .
\item[(ii)] $|a_\infty(\lambda)|^2 + |b_\infty(\lambda)|^2  = 1, \qquad \text{\upshape Im} \, \lambda = 0, \quad \lambda \in \bar{D}_5$.
\end{enumerate}
\end{proposition}
\proofbegin
Property (i) follows from the discussion in subsection \ref{eigeninftysubsec}.
Property (ii) follows from (\ref{sinftydet1}).
\proofend

\section{The Riemann-Hilbert problem}\label{RHsec} 
\begin{figure}
\begin{center}
  \includegraphics[width=.5\textwidth]{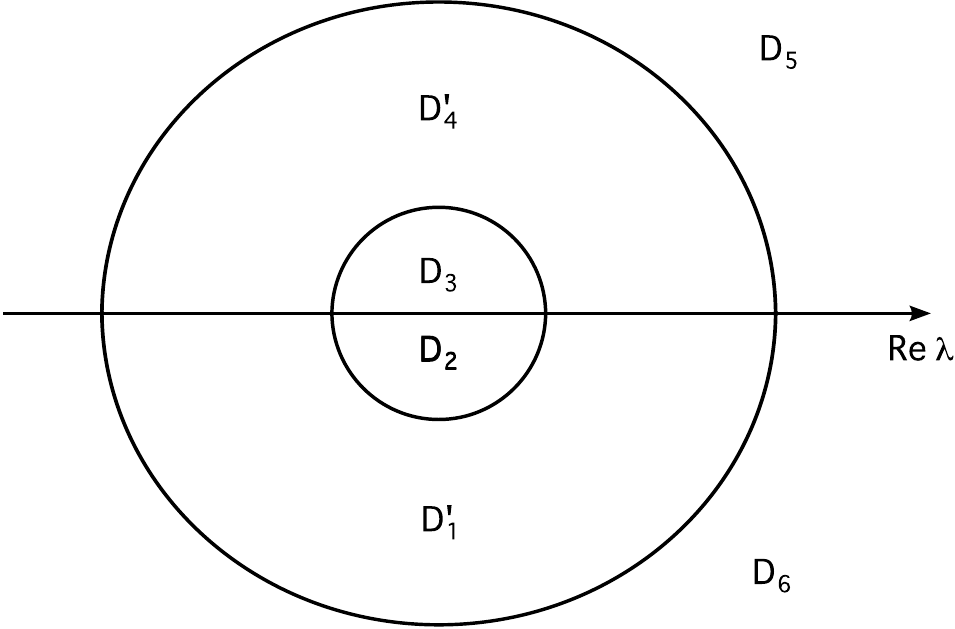} \\
     \begin{figuretext}\label{RHcontour.pdf}
        The contour for the Riemann-Hilbert problem in the complex $\lambda$-plane.
     \end{figuretext}
     \end{center}
\end{figure}

\nequation
In this section we use the eigenfunctions $\{\mu_j\}_1^3$ and $\{\Phi_j\}_2^3$ to formulate a RH problem for a $2 \times 2$-matrix valued function with jump contour shown in Figure \ref{RHcontour.pdf}. We will first formulate a RH problem for a $2 \times 2$-matrix valued function $\tilde{M}(x,t,\lambda)$, whose form is suggested by the methodology of \citet{F1997}. However, it turns out that the jump matrix for this RH problem depends on the function $p(x,t)$ which occurs in the Lax pair (\ref{Philax}). The function $p(x,t)$ is unknown from the point of view of the inverse problem, and thus the solution is not yet effective. We will overcome this problem by introducing new variables $(y, \eta)$ and formulating a modified RH problem for a $2 \times 2$-matrix valued function $M(y, \eta,\lambda)$, whose jump condition is given explicitly in terms of $y$, $\eta$, and $\lambda$. The solution $u(x,t)$ of (\ref{gsglab}) can be recovered in parametric from the asymptotics of $M(y,\eta, \lambda)$. 
A similar reparametrization of the RH problem occurs also in the analysis of other equations such as the Camassa-Holm equation and equation (\ref{gsg}), although in those cases only one of the variables $(x,t)$ has to be reparametrized, cf. \citet{BS2008, LFgsg}.

\subsection{RH problem for $\tilde{M}(x,t,\lambda)$}
We seek a bounded and sectionally analytic $2 \times 2$-matrix valued function $\tilde{M}(x,t,\lambda)$, which satisfies a jump condition of the form
\begin{align}\label{tildeMMJ}
& \tilde{M}_-(x,t,\lambda) = \tilde{M}_+(x,t, \lambda)\tilde{J}(x,t,\lambda), \qquad \lambda \in \bar{D}_+ \cap \bar{D}_-,
	\\
& \tilde{M} = \begin{cases} \tilde{M}_+, \qquad \lambda \in \bar{D}_+, \\
\tilde{M}_-, \qquad \lambda \in \bar{D}_-,
\end{cases}
\end{align}
where $\tilde{J}(x,t,\lambda)$ is a $2 \times 2$-matrix valued `jump matrix' and
$$D_+ = D_1' \cup D_3 \cup D_5, \qquad D_- = D_2 \cup D_4' \cup D_6.$$
Since the $\mu_j$'s and the $\Phi_j$'s are well-behaved near $\lambda = 0$ and $\lambda = \infty$, respectively, we define $\tilde{M}$ in terms of the $\mu_j$'s in the regions $D_1'$, $D_2$, $D_3$, and $D_4'$, and in terms of $\Phi_2$ and $\Phi_3$ in the regions $D_5$ and $D_6$.
The methodology of \citet{F1997} suggests making the following ansatz for $\tilde{M}$:
\begin{equation} \label{Mtildedef}
\tilde{M}(x,t, \lambda) =
\left\{ \begin{array}{ll}
g(x,t) \left(\frac{\mu_2^{(1)}}{a(\lambda)}, \mu_3^{(12)}\right), & \qquad \lambda \in \bar{D}_1',
 	\\
g(x,t) \left(\frac{\mu_1^{(2)}}{d(\lambda)}, \mu_3^{(12)}\right), & \qquad\lambda \in \bar{D}_2,
		\\
g(x,t) \left(\mu_3^{(34)}, \frac{\mu_1^{(3)}}{\overline{d(\bar{\lambda})}}\right), & \qquad \lambda \in \bar{D}_3,
	\\
g(x,t)  \left(\mu_3^{(34)}, \frac{\mu_2^{(4)}}{\overline{a(\bar{\lambda})}}\right), & \qquad \lambda \in \bar{D}_4',
		\\
 \left(\frac{\Phi_2^{(5)}}{a_\infty(\lambda)}, \Phi_3^{(5)}\right), & \qquad \lambda \in \bar{D}_5 
	\\
\left(\Phi_3^{(6)}, \frac{\Phi_2^{(6)}}{\overline{a_\infty(\bar{\lambda})}}\right), & \qquad \lambda \in \bar{D}_6,
\end{array} \right.
\end{equation}
where
\begin{equation}\label{ddef}
  d(\lambda) =a(\lambda)\overline{A(\bar{\lambda})} + b(\lambda) \overline{B(\bar{\lambda})}, \qquad \lambda \in \bar{D}_2.
\end{equation}
The definition of $\tilde{M}$ in $D_1' \cup D_2 \cup D_3 \cup D_4'$, which involves the $\mu_j$'s, includes the prefactor $g(x,t)$. This prefactor is suggested by the relationship (\ref{phigpsi}) between eigenfunctions of the Lax pairs (\ref{mulax}) and (\ref{Philax}), and its inclusion implies that there exists a jump matrix $\tilde{J}$ such that $\tilde{M}_+$ and $\tilde{M}_-$ are related as in (\ref{tildeMMJ}).
We introduce the following notation:
\begin{equation}\label{J1234567def}
\tilde{J}(x,t,\lambda) = \left\{ \begin{array}{ll}
\tilde{J}_1, & \qquad \lambda \in \bar{D}_1' \cap \bar{D}_2,  \\
\tilde{J}_2 = \tilde{J}_3 \tilde{J}_4^{-1} \tilde{J}_1, & \qquad \lambda \in \bar{D}_2 \cap \bar{D}_3, \\
\tilde{J}_3, & \qquad \lambda \in \bar{D}_3 \cap \bar{D}_4', \\
\tilde{J}_4, & \qquad \lambda \in \bar{D}_4' \cap \bar{D}_1', \\
\tilde{J}_5, & \qquad \lambda \in \bar{D}_4' \cap \bar{D}_5, \\
\tilde{J}_6, & \qquad \lambda \in \bar{D}_5 \cap \bar{D}_6, \\
\tilde{J}_7 = \tilde{J}_4 \tilde{J}_5^{-1} \tilde{J}_6, & \qquad \lambda \in \bar{D}_1' \cap \bar{D}_6.
\end{array} \right.
\end{equation}
The jump matrices $\{\tilde{J}_n\}_1^7$ can be determined from the various relations between the eigenfunctions.
Indeed, algebraic manipulation of the equations (\ref{sSdef}) leads to expressions for the jump matrices $\{\tilde{J}_n\}_1^4$ in terms of the spectral functions $s(\lambda)$ and $S(\lambda)$. Similarly, algebraic manipulation of equation (\ref{sinftydef}) leads to an expression for the jump matrix $\tilde{J}_6$ in terms of the spectral function $s_\infty(\lambda)$. 
To find an expression for the jump matrix $\tilde{J}_5$, we note that the relations (\ref{phigpsi}), (\ref{psimurel}), and (\ref{phiPhirel}) imply that two solutions $\mu$ and $\Phi$ of (\ref{mulax}) and (\ref{Philax}), respectively, satisfy a relation of the form 
\begin{equation}\label{muPhirel}
 g(x,t) \mu(x,t,\lambda) = \Phi(x,t,\lambda) e^{-i \theta_\infty(x,t,\lambda) \sigma_3} C(\lambda) e^{i\theta(x,t,\lambda)\sigma_3},
\end{equation}
where $C(\lambda)$ is a $2\times 2$-matrix independent of $(x, t)$ and the functions $\theta(x,t,\lambda)$ and $\theta_\infty(x,t,\lambda)$ are defined by
\begin{equation}\label{thetadef}
\theta(x,t,\lambda) = \frac{x}{8\lambda} - (\lambda + \frac{1}{8\lambda}) t, \qquad \theta_\infty(x,t,\lambda) = \lambda p(x,t) + \frac{t - x}{8\lambda}.
\end{equation}
In the particular case of $\mu = \mu_3$ and $\Phi = \Phi_2$, equation (\ref{muPhirel}) holds with $C(\lambda) = g(0,0)s(\lambda)$, i.e.
\begin{equation}\label{mu3Phi2rel}
  g(x,t)\mu_3(x,t,\lambda) = \Phi_2(x,t,\lambda) e^{-i \theta_\infty(x,t,\lambda) \sigma_3} g(0,0)s(\lambda) e^{i\theta(x,t,\lambda)\sigma_3}.
\end{equation}
Equation (\ref{mu3Phi2rel}) together with (\ref{sdef}) and (\ref{sinftydef}) provide the required relations between the $\mu_j$'s and the $\Phi_j$'s needed for determining $\tilde{J}_5$. 
In summary, we arrive at the following expressions for the $\tilde{J}_n$'s:
\begin{align*}
&\tilde{J}_1 = e^{-i\theta\hat{\sigma}_3}J_1^0, \qquad
\tilde{J}_3 = e^{-i\theta\hat{\sigma}_3}J_3^0, \qquad
\tilde{J}_4 = e^{-i\theta\hat{\sigma}_3}J_4^0,
	\\
& \tilde{J}_5 = e^{-i\theta_\infty \sigma_3} J_5^0 e^{i\theta \sigma_3}, \qquad
\tilde{J}_6 = e^{-i\theta_\infty \hat{\sigma}_3}J_6^0,
\end{align*}
where 
\begin{align} \label{J0def} 
& J_1^0 = \begin{pmatrix} 1	&	0 	\\
\Gamma(\lambda)	&	1 \end{pmatrix}, \qquad 
J_4^0 =\begin{pmatrix} 1	&	-\frac{b(\lambda)}{\overline{a(\bar{\lambda})}}	\\
- \frac{\overline{b(\bar{\lambda})}}{a(\lambda)}	&	\frac{1}{a(\lambda)\overline{a(\bar{\lambda})}} \end{pmatrix}, 
\qquad  J_3^0 =  \begin{pmatrix} 1	&	\overline{\Gamma(\bar{\lambda})} 	\\
0	&	1 \end{pmatrix},
	\\ \nonumber
& J_5^0 = \begin{pmatrix} (s_\infty^{-1}(\lambda) g(0,0) s(\lambda))_{11} 
& \frac{(s_\infty^{-1}(\lambda)g(0,0))_{12}}{\overline{a(\bar{\lambda})}} \\
\frac{(g(0,0)s(\lambda))_{21}}{a_\infty(\lambda)} & \frac{g(0,0)_{22}}{a_\infty(\lambda) \overline{a(\bar{\lambda})}} \end{pmatrix}, 	
\qquad J_6^0 = \begin{pmatrix} 1	&	-\frac{b_\infty(\lambda)}{\overline{a_\infty(\bar{\lambda})}}	\\
- \frac{\overline{b_\infty(\bar{\lambda})}}{a_\infty(\lambda)}	&	\frac{1}{a_\infty(\lambda)\overline{a_\infty(\bar{\lambda})}} \end{pmatrix},
\end{align}
and $\Gamma(\lambda)$ is defined by
\begin{equation}\label{Gammadef}
   \Gamma(\lambda) = -\frac{\overline{B(\bar{\lambda})}}{a(\lambda)d(\lambda)}, \qquad \lambda \in \bar{D}_2.
\end{equation}

\subsection{RH problem for $M(y, \eta, \lambda)$}
In the previous subsection we formulated a RH problem for $\tilde{M}(x,t,\lambda)$ in the Riemann sphere of the spectral parameter $\lambda$. However, as noted above, this RH problem does not provide the solution of our initial-boundary value problem. Indeed, the jump matrices $\{\tilde{J}_n\}_5^7$ involve $\theta_\infty$. The occurence of the function $p(x,t)$ in $\theta_\infty$ implies that the RH problem cannot be formulated in terms of the initial and boundary data alone. To overcome this problem we make two important changes in the formulation of the RH problem:
(a) We modify the jump matrix by adding appropriate exponential factors to the definition (\ref{Mtildedef});
(b) We introduce new variables $(y, \eta)$ by
\begin{equation}\label{yeta}
  (x,t) \mapsto (y, \eta), \qquad  y = p(x,t), \qquad \eta = \frac{1}{2}(x - t),
\end{equation}
where $p(x,t)$ was defined in (\ref{pdef}). The jump matrix of the modified RH problem is explicitly given in terms of $(y, \eta, \lambda)$ and can thus be formulated in terms of the initial and boundary data alone.

\begin{figure}
\begin{center}
  \includegraphics[width=.5\textwidth]{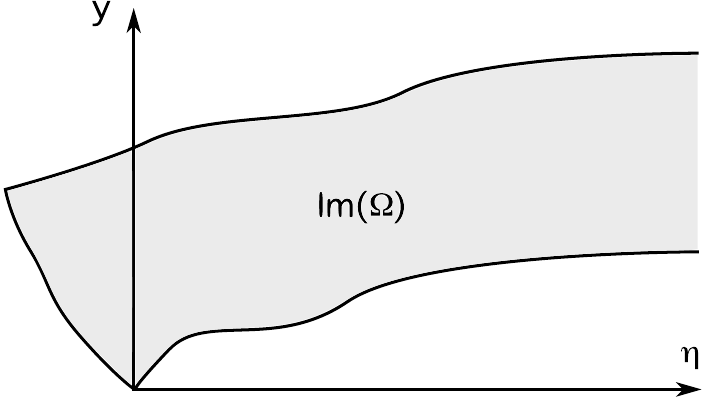} \\
     \begin{figuretext}\label{yetadomain.pdf}
        The image in the $(y, \eta)$-plane of the half-line domain $\Omega = \{0 \leq x < \infty, 0 \leq t \leq T\}$ under the map (\ref{yeta}).
     \end{figuretext}
     \end{center}
\end{figure}

Define the $2Ê\times 2$-matrix valued function $M(y, \eta, \lambda)$ by
\begin{align}  \label{Mdef}
M(y,\eta, \lambda) =
\left\{ \begin{array}{ll}
g \left(\frac{\mu_2^{(1)}}{a(\lambda)}, \mu_3^{(12)}\right)e^{i\lambda(y + t)\sigma_3}, & \qquad\lambda \in \bar{D}_1',
 	\\
g \left(\frac{\mu_1^{(2)}}{d(\lambda)}, \mu_3^{(12)}\right)e^{i\lambda(y + t)\sigma_3}, & \qquad\lambda \in \bar{D}_2,
		\\
g \left(\mu_3^{(34)}, \frac{\mu_1^{(3)}}{\overline{d(\bar{\lambda})}}\right)e^{i\lambda(y + t)\sigma_3}, & \qquad\lambda \in \bar{D}_3,
	\\
g  \left(\mu_3^{(34)}, \frac{\mu_2^{(4)}}{\overline{a(\bar{\lambda})}}\right)e^{i\lambda(y + t)\sigma_3}, & \qquad \lambda \in \bar{D}_4',
		\\
 \left(\frac{\Phi_2^{(5)}}{a_\infty(\lambda)}, \Phi_3^{(5)}\right)e^{i\frac{\eta}{2\lambda} \sigma_3}, & \qquad \lambda \in \bar{D}_5,
	\\
\left(\Phi_3^{(6)}, \frac{\Phi_2^{(6)}}{\overline{a_\infty(\bar{\lambda})}}\right)e^{i\frac{\eta}{2\lambda} \sigma_3}, & \qquad \lambda \in \bar{D}_6,
\end{array} \right.
\end{align}
The map (\ref{yeta}) is a bijection from $\Omega = \{0 \leq x < \infty, 0 \leq t \leq T\}$ to a subset of the $(y,\eta)$-plane delimited by the image of $\partial\Omega$ under (\ref{yeta}), see Figure \ref{yetadomain.pdf}. The image of $\partial \Omega$ consists of three pieces given by
\begin{align}
& \left\{(y, \eta) = \frac{1}{2}\left(\int_0^x (1 - \cos(u(x',0)))\sqrt{m(x',0)} dx', x\right) \biggl| x \geq 0\right\},
	\\
& \left\{(y, \eta) = \frac{1}{2}\left(\int_0^t (1 + \cos(u(0, t')))\sqrt{m(0,t')} dt', -t\right) \biggl| 0 \leq t \leq T \right\},
	\\
& \left\{(y, \eta) = \frac{1}{2}\left(p(0, T) + \int_0^x (1 - \cos(u(x',T)))\sqrt{m(x',T)} dx', x-T\right) \biggl| x \geq 0\right\}.
\end{align}
In particular, the map (\ref{yeta}) is everywhere nonsingular:
$$\begin{vmatrix} \frac{\partial y}{\partial x} & \frac{\partial y}{\partial t} \\
\frac{\partial \eta}{\partial x} & \frac{\partial \eta}{\partial t} \end{vmatrix}
= \begin{vmatrix} \frac{1}{2}(1 - \cos{u})\sqrt{m} & \frac{1}{2}(1 + \cos{u})\sqrt{m}  \\
\frac{1}{2}  & - \frac{1}{2}  \end{vmatrix}
= - \frac{1}{2}\sqrt{m} < 0.$$
The expressions on the right-hand side of (\ref{Mdef}) should be understood as being evaluated at the point $(x,t)$ corresponding to $(y, \eta)$ under (\ref{yeta}). 

The form of the exponential factors $e^{i\lambda(y + t)\sigma_3}$ and $e^{i\frac{\eta}{2\lambda} \sigma_3}$ on the right-hand side of (\ref{Mdef}) is motivated by the fact that these exponential factors are analytic near $\lambda = 0$ and $\lambda = \infty$, respectively, and by the relations
\begin{equation}\label{thetathetainfty}
e^{i\lambda(y + t)\sigma_3}e^{i\theta\sigma_3} = e^{i(\lambda y + \frac{\eta}{4\lambda})\sigma_3}, \qquad e^{i\frac{\eta}{2\lambda}\sigma_3}e^{i\theta_\infty \sigma_3} = e^{i(\lambda y + \frac{\eta}{4\lambda})\sigma_3}.
\end{equation}
The relations (\ref{thetathetainfty}) imply that $M$ satisfies the jump conditon
\begin{align}\label{MMJ}
 &M_-(y, \eta, \lambda) = M_+(x,t, \lambda)J(y, \eta, \lambda), \qquad \lambda \in \bar{D}_+ \cap \bar{D}_-;
	\\
&M = \begin{cases} M_+, \qquad \lambda \in \bar{D}_+, \\
M_-, \qquad \lambda \in \bar{D}_-,
\end{cases}
\end{align}
where the jump matrix $J(y,\eta, \lambda)$ is given by 
\begin{align}\label{Jdef}
 &J_n(y,\eta, \lambda) = e^{-i (\lambda y + \frac{\eta}{4\lambda}) \hat{\sigma}_3}J_n^0(\lambda), \qquad n = 1, \dots, 7;
 	\\ \nonumber
& J = \left\{ \begin{array}{ll}
J_1 & \qquad \lambda \in \bar{D}_1' \cap \bar{D}_2,  \\
J_2 = J_3 J_4^{-1} J_1 & \qquad \lambda \in \bar{D}_2 \cap \bar{D}_3, \\
J_3 & \qquad \lambda \in \bar{D}_3 \cap \bar{D}_4', \\
J_4 & \qquad \lambda \in \bar{D}_4' \cap \bar{D}_1', \\
J_5 & \qquad \lambda \in \bar{D}_4' \cap \bar{D}_5, \\
J_6 & \qquad \lambda \in \bar{D}_5 \cap \bar{D}_6, \\
J_7 = J_4 J_5^{-1} J_6 & \qquad \lambda \in \bar{D}_1' \cap \bar{D}_6.
\end{array} \right.
\end{align}
We can now prove the following theorem.

\begin{theorem}\label{RHtheorem}
Let $T \leq \infty$ and let $u_0(x)$, $u_1(x)$, $0 \leq x \leq \infty$, and $g_0(t)$,  $g_1(t)$, $0 \leq t \leq T$, be given functions. Suppose that there exists a solution $u(x,t)$ of equation (\ref{gsglab}) in the domain $\{0 \leq x < \infty, 0 \leq t \leq T\}$, which satisfies the initial conditions
$$u(x, 0) = u_0(x), \quad u_t(x,0) = u_1(x), \qquad 0 \leq x < \infty,$$
and the boundary conditions
$$u(0,t) = g_0(t), \quad u_x(0,t) = g_1(t), \qquad 0 \leq t \leq T.$$
Moreover, suppose that $\cos(u(x,t)) - 1$ has sufficient smoothness and decay as $x \to \infty$ and that $\cos(g_0(t)) \neq -1$ for $0 \leq t \leq T$.\footnote{If $T = \infty$, we also assume that $\cos(u(0,t)) - 1$ has decay as $t \to \infty$.}
Then $u(x,t)$ can be reconstructed from the initial and boundary values $\{u_0(x), u_1(x), g_0(t), g_1(t)\}$ as follows.

Define the functions $m(x, 0)$ and $m(0, t)$ by
$$m(x, 0) = 1 + (u_{0x}(x) + u_1(x))^2, \qquad
m(0, t) = 1 + (g_1(t) + g_{0t}(t))^2.$$
Define the functions $p(x, 0)$ and $p(0, t)$ by
\begin{align*}
& p(x, 0) = \frac{1}{2} \int_0^x (1 - \cos(u_0(x')))\sqrt{m(x', 0)}dx', \qquad 0 \leq x < \infty,
	\\
& p(0, t) = \frac{1}{2} \int_0^t(1 + \cos(g_0(t')))\sqrt{m(0, t')}dt', \qquad 0 \leq t \leq T.
\end{align*}
Define $\mu_3(x, 0, \lambda)$ and $\Phi_3(x, 0, \lambda)$ in terms of $u_0(x)$, $u_1(x)$, and $p(x,0)$ via the Volterra linear integral equations (\ref{mujdef}) and (\ref{Phijdef}) evaluated at $t = 0$. Define the spectral functions $a(\lambda)$, $b(\lambda)$, $a_\infty(\lambda)$, $b_\infty(\lambda)$ by equations (\ref{sSmurel}) and (\ref{sinftyPhirel}).
Similarly, define $\mu_1(x, 0, \lambda)$ in terms of $g_0(t)$ and $g_1(t)$ via the Volterra linear integral equation (\ref{mujdef}) evaluated at $x = 0$. Define the spectral functions $A(\lambda)$, $B(\lambda)$ by equation (\ref{sSmurel}).

Let $R > 0$ be such that
\begin{equation*}
  R^2 \geq \sup_{0 \leq t', t \leq T} \frac{t - t'}{4\int_{t'}^{t} (1 + \cos(g_0(\tau))\sqrt{m(0,\tau)} d\tau},
\end{equation*}
and define the sets $\{D_n\}_1^6$ by (\ref{D1234def}) and (\ref{D56def}). Define $D_1'$ and $D_4'$ by (\ref{D12primedef}).
Let 
$$D_+ = D_1' \cup D_3 \cup D_5, \qquad D_- = D_2 \cup D_4' \cup D_6.$$
Define $d(\lambda)$ and $\Gamma(\lambda)$ by (\ref{ddef}) and (\ref{Gammadef}). Assume that the possible zeros $\{k_j\}_{1}^{N}$ of $a(\lambda)$ and $\{\lambda_j\}_{1}^{\Lambda}$ of $d(\lambda)$ satisfy
\begin{itemize}
\item $a(k)$ has $N$ simple zeros $\{k_j\}_{j = 1}^{N}$ such that $k_j \in D_1'$, $j = 1, \dots, n_1$, and $k_j \in D_2$, $j = n_1+1, \dots, N$. 

\item $d(k)$ has $\Lambda$ simple zeros $\{\lambda_j\}_1^{\Lambda}$, such that 
$\lambda_j \in D_2$, $j = 1, \dots, \Lambda.$ 

\item None of the zeros of $a(k)$ coincides with a zero of $d(k)$.
\end{itemize}

Then the solution $u(x,t)$ of equation (\ref{gsglab}) is given in parametric form in terms of two real parameters $y, \eta$ such that
\begin{equation}\label{etayineqs}
 \eta  \geq -\frac{T}{2}; \qquad y \geq p(2\eta, 0) \quad \text{\upshape for} \quad \eta \geq 0; \qquad y \geq p(0, 2|\eta|)\quad \text{\upshape for} \quad  -\frac{T}{2} \leq \eta \leq 0,
\end{equation}
by
\begin{subequations}\label{recoveru}
\begin{align}\label{recoverua}
  u\left(\xi(y,\eta) + \eta, \xi(y,\eta) - \eta\right) 
  = \begin{cases}
   2 \text{\upshape Im}\left(\int_{p(2\eta,0)}^y \alpha(y',\eta) dy'\right) + u_0(2\eta), \qquad \eta \geq 0, \\
   2 \text{\upshape Im}\left(\int_{p(0,2|\eta|)}^y \alpha(y',\eta) dy'\right) + g_0(2|\eta|), \qquad -\frac{T}{2} \leq \eta \leq 0,
   \end{cases}
\end{align}
where $\xi(y,\eta)$ is defined by
\begin{align}\label{xiyetadef}
 \xi(y,\eta) = \begin{cases}
  \int_{p(2\eta,0)}^y \sqrt{1 - 4\text{\upshape Im}(\alpha(y',\eta))^2} dy' + \eta, \qquad \eta \geq 0,
	\\
  \int_{p(0, 2|\eta|)}^y \sqrt{1 - 4\text{\upshape Im}(\alpha(y',\eta))^2} dy' + |\eta|, \qquad -\frac{T}{2} \leq \eta \leq 0,
\end{cases}	
\end{align}
the function $\alpha(y,\eta)$ is the unique solution of the Ricatti equation
\begin{align}\label{ricatti}
 \alpha_y =  \alpha^2 - 4 i  \left(\lim_{\substack{\lambda \to \infty \\ \text{\upshape Im}\,\lambda > 0}} \lambda M_{12}(y,\eta,\lambda)\right)\alpha - \frac{1}{4},
\end{align}
together with the initial conditions
\begin{align}\label{ricattiinitial}
& \alpha\left(p(2\eta,0), \eta\right) = -\frac{1}{2}e^{-i \arcsin\left(\frac{u_{0x}(2\eta) + u_1(2\eta)}{\sqrt{1 + (u_{0x}(2\eta) + u_1(2\eta))^2}}\right)}, \qquad \eta \geq 0,
 	\\ \nonumber
& \alpha\left(p(0,2|\eta|), \eta\right) = -\frac{1}{2}e^{-i \arcsin\left(\frac{g_{1}(2|\eta|) + g_{0t} (2|\eta|)}{\sqrt{1 + (g_1(2|\eta|) + g_{0t}(2|\eta|))^2}}\right)}, \qquad -\frac{T}{2} \leq \eta \leq 0,	
 \end{align}
\end{subequations}
and $M(y, \eta, \lambda)$ is the unique solution of the following $2\times 2$-matrix RH problem:
\begin{itemize}
\item $M$ is meromorphic away from the contour $\bar{D}_+ \cap \bar{D}_-$.

\item The possible poles of the first column of $M$ occur at $\lambda = k_j$, $j = 1, \dots, n_1$, and $\lambda = \lambda_j$, $j = 1, \dots, \Lambda$. The possible poles of the second column of $M$ occur at $\lambda = \bar{k}_j$, $j = 1, \dots, n_1$, and $\lambda = \bar{\lambda}_j$, $j = 1, \dots, \Lambda$. The associated residues satisfy the following relations:
\begin{subequations}\label{residue}
\begin{align}\label{residue1}
\underset{k_j}{\text{\upshape Res}} [M(y,\eta,\lambda)]_1 =& \frac{1}{\dot{a}(k_j)b(k_j)} e^{2i(k_j y + \frac{\eta}{4k_j})\sigma_3} [M(y, \eta,k_j)]_2, \qquad j = 1, \dots, n_1,
		\\\label{residue2}
\underset{\bar{k}_j}{\text{\upshape Res}} [M(y,\eta,\lambda)]_2 =& \frac{-1}{\overline{\dot{a}(k_j)b(k_j)}} e^{-2i(\bar{k}_j y + \frac{\eta}{4\bar{k}_j})\sigma_3}[M(y,\eta,\bar{k}_j)]_1, \qquad j = 1, \dots, n_1,
	\\ \label{residue3}
\underset{\lambda_j}{\text{\upshape Res}} [M(y,\eta,\lambda)]_1 = &
\underset{\lambda_j}{\text{\upshape Res}} \, \Gamma(k) \, e^{2i(\lambda_j y + \frac{\eta}{4\lambda_j})\sigma_3}[M(y,\eta,\lambda_j)]_2, \qquad j = 1, \dots, \Lambda,
	\\\label{residue4}
\underset{\bar{\lambda}_j}{\text{\upshape Res}} [M(y,\eta,\lambda)]_2 =& 
- \underset{\bar{\lambda}_j}{\text{\upshape Res}} \, \overline{\Gamma(\bar{k})} \,
e^{-2i(\bar{\lambda}_j y + \frac{\eta}{4\bar{\lambda}_j})\sigma_3}
 [M(y,\eta,\bar{\lambda}_j)]_1, \qquad j = 1, \dots, \Lambda.
\end{align}
\end{subequations}

\item $M$ satisfies the jump condition
$$M_-(y,\eta,\lambda) = M_+(y,\eta,\lambda)J(y,\eta,\lambda), \qquad \lambda \in \bar{D}_+ \cap \bar{D}_-,$$
where $M$ is $M_-$ for $\lambda \in D_-$, $M$ is $M_+$ for $\lambda \in D_+$, and $J$ is defined by equations (\ref{J0def}) and (\ref{Jdef}).

\item $M(y,\eta,\lambda) = I + O(\lambda), \qquad \lambda \to 0.$

\item $M(y,\eta,\lambda) = I + O\left(\frac{1}{\lambda}\right), \qquad \lambda \to \infty.$
\end{itemize}
\end{theorem}
\proofbegin
In the case when $a(\lambda)$ and $d(\lambda)$ have no zeros, the unique solvability is a consequence of the existence of a vanishing lemma.
If $a(\lambda)$ and $d(\lambda)$ have zeros, the singular RH problem can be mapped to a regular one coupled with a system of algebraic equations, see \citet{F-I}.

The residue conditions (\ref{residue}) can be proved as follows. The general approach of \citet{Fbook} implies that $\tilde{M}$ as defined in (\ref{Mtildedef}) satisfies the residue condition
$$\underset{k_j}{\text{Res}} [\tilde{M}(x,t,\lambda)]_1 = \frac{1}{\dot{a}(k_j)b(k_j)} e^{2i\theta(x,t, k_j)} [\tilde{M}(x,t,k_j)]_2, \qquad j = 1, \dots, n_1,$$
where $\theta(x,t,\lambda)$ is given by (\ref{thetadef}). The definition (\ref{Mdef}) of $M$ and the relations (\ref{thetathetainfty}) imply that $M$ satisfies (\ref{residue1}). The other residue conditions follow similarly from the corresponding residue conditions for $\tilde{M}$.

In order to prove (\ref{recoveru}), we note that the change of variables (\ref{yeta}) implies that
$$\partial_x = \frac{1}{2}(1 - \cos{u})\sqrt{m} \partial_y + \frac{1}{2}\partial_\eta, \qquad \partial_t = \frac{1}{2}(1 +\cos{u})\sqrt{m} \partial_y - \frac{1}{2}\partial_\eta.$$
In particular, 
\begin{equation}\label{partialxty}  
  \partial_x + \partial_t = \sqrt{m}\partial_y.
\end{equation}
Moreover, because
$$m = 1 + (u_x + u_t)^2 = 1 + m u_y^2,$$
we find that
$$ m = \frac{1}{1 - u_y^2}.$$
Thus, addition of the two equations in (\ref{Philax}) together with (\ref{yeta}) yields
\begin{equation}\label{Phiypart}  
  \Phi_y + i\lambda [\sigma_3, \Phi] = \frac{i}{2}\left(u_y + \frac{u_{yy}}{\sqrt{1 - u_y^2}}\right)\sigma_1\Phi.
\end{equation}
Consider the particular solution of this equation given by
$$\Phi = \left(\frac{\Phi_2^{(5)}}{a_\infty(\lambda)}, \Phi_3^{(5)}\right).$$
Equation (\ref{Phiypart}) implies that $\Phi$ admits an expansion of the form
\begin{equation}\label{Phiexpansion} 
 \Phi(x,t,\lambda) = I + \frac{\Phi^{(1)}(x,t)}{\lambda} + \frac{\Phi^{(2)}(x,t)}{\lambda^2} + O\left(\frac{1}{\lambda^3}\right), \qquad \lambda \to \infty,\quad \lambda \in D_5,
\end{equation}
where $\Phi^{(1)}(x,t)$ and $\Phi^{(2)}(x,t)$ are independent of $\lambda$. Substituting this expansion into (\ref{Phiypart}) we find by considering the terms of $O(1)$ that
$$   4 \Phi^{(1)}_{12}(x,t) = u_y + \frac{u_{yy}}{\sqrt{1 - u_y^2}}.$$
Thus, by construction of the RH problem,
\begin{equation}\label{4limM12}  
4 \lim_{\substack{\lambda \to \infty \\ \text{\upshape Im}\,\lambda > 0}} \lambda M_{12}(y,\eta,\lambda)
= u_y + \frac{u_{yy}}{\sqrt{1 - u_y^2}}.
\end{equation}
In view of the inquality
$$-1 \leq u_y = \frac{u_x + u_t}{\sqrt{1 + (u_x + u_t)^2}} \leq 1,$$
we may define two functions $Q(y, \eta)$ and $\alpha(y,\eta)$ by
\begin{equation}\label{Qalphadef}  
Q = -u - \arcsin(u_y), \qquad \alpha = -\frac{1}{2} e^{-i\arcsin(u_y)}.
\end{equation}
These definitions imply that 
\begin{equation}\label{uyQy}  
  u_y = 2\text{Im}(\alpha), \qquad Q_y = -\left(u_y + \frac{u_{yy}}{\sqrt{1 - u_y^2}}\right),
\end{equation}
and that $\alpha$ satisfies the Ricatti equation
\begin{equation}\label{ricatti2}  
  \alpha_y = \alpha^2 + iQ_y \alpha - \frac{1}{4}.
\end{equation}
Equation (\ref{ricatti}) follows from (\ref{4limM12}), (\ref{uyQy}) and (\ref{ricatti2}). 

Equation (\ref{Phiexpansion}) implies that $M = I + O(1/\lambda)$ as $\lambda \to \infty$ in $D_5$. A similar argument shows that $M = I + O(1/\lambda)$ as $\lambda \to \infty$ also in $D_6$.

In order to prove that $\alpha$ satisfies the initial conditions (\ref{ricattiinitial}) we note that the initial half-line $\{(x, t) = (2s, 0) | s > 0\}$ is mapped by (\ref{yeta}) to the set $\{(y, \eta) = (p(2s, 0), s) | s > 0\}$. Together with the definition (\ref{Qalphadef}) of $\alpha$ this leads to
$$\alpha(p(2\eta, 0), \eta) = -\frac{1}{2}e^{-i\arcsin(u_y)}\bigl|_{\substack{x = 2\eta \\ t = 0}}, \qquad \eta \geq 0,$$
which yields (\ref{ricattiinitial}) for $\eta \geq 0$. The proof when $-T/2 \leq \eta \leq 0$ is similar.

Finally, equations (\ref{recoverua}) and (\ref{xiyetadef}) can be derived by considering the map (\ref{yeta}) as the composition of the change of variables (\ref{lighttolab}) with the map
\begin{equation}\label{xietayeta} 
  (\xi, \eta) \mapsto (y, \eta) = \left(\int_{|\eta|}^\xi \sqrt{m} d\xi' + p(|\eta| + \eta, |\eta| - \eta), \eta\right).
\end{equation}
Thus, the variables $(y, \eta) \in \text{Im}(\Omega)$ satisfy (\ref{etayineqs}).
Using that the map (\ref{xietayeta}) admits the inverse
$$(y, \eta) \mapsto (\xi, \eta) =  \left(\int_{p (|\eta| + \eta, |\eta| - \eta)}^y m^{-1/2} dy' + |\eta|, \eta\right),$$
together with the expression (\ref{uyQy}) for $u_y$ in terms of $\text{Im}\,\alpha$, we find (\ref{recoverua}) and (\ref{xiyetadef}).

\proofend

\bibliography{gsghalfline4}

\end{document}